\documentclass[journal=jacsat,manuscript=article]{achemso}

\usepackage[utf8]{inputenc}
\usepackage[T1]{fontenc}
\usepackage{hyperref}
\usepackage{url}
\usepackage{booktabs}
\usepackage{amsfonts}
\usepackage{amsmath}
\usepackage{nicefrac}
\usepackage{microtype}
\usepackage[version=3]{mhchem} 
\usepackage{subcaption} 
\usepackage{graphicx}
\usepackage{listings}
\usepackage{xcolor}
\usepackage{textcomp}   

\author{Emmanuel George\textsuperscript{1}}
\affiliation{
  Department of Mechanical Engineering, Carnegie Mellon University, Pittsburgh,
  PA, USA
}

\author{Christopher Keefe\textsuperscript{2}}
\affiliation{
  Department of Mechanical Engineering, Carnegie Mellon University, Pittsburgh,
  PA, USA
}

\author{Peter Pak\textsuperscript{3}}
\affiliation{
  Department of Mechanical Engineering, Carnegie Mellon University, Pittsburgh,
  PA, USA
}

\author{Amir Barati Farimani}
\email{barati@cmu.edu}
\affiliation{
  Department of Mechanical Engineering, Carnegie Mellon University, Pittsburgh,
  PA, USA
}

\title[]{
    AgentsCAD: Automated Design for Manufacturing of FDM Parts via Multi-Agent LLM
    Reasoning and Geometric Feature Recognition
}
\abbreviations{FDM, DFM, LLM, VLM, B-Rep, STEP}

\SectionNumbersOn

\begin{document}

\begin{abstract}
Parts manufactured with Fused Deposition Modeling (FDM) often require Design for
Additive Manufacturing (DFAM) modifications to ensure printability, structural
integrity, and reduced post-processing. Current slicers identify defects such as
steep overhangs but are unable to modify the underlying geometry. This work
presents \textit{AgentsCAD}, a multi-agent system that bridges raw
boundary-representation (B-Rep) geometry and Large Language Model (LLM)
reasoning to automate targeted DFM. The workflow begins by parsing a
\texttt{STEP} file. The agentic system detects overhangs above a $45$\textdegree
threshold, constructs a face-adjacency topology graph, and optionally injects
semantic feature labels from a GraphSAGE model trained on MFCAD++ (59{,}665
parts), before dispatching a Claude~Sonnet design-reasoning agent that
recommends reorientations, fillets, chamfers, and similar modifications. A
GPT-4o vision-language verifier inspects rendered views to confirm geometric
integrity. Outputs include a modified \texttt{STEP} file and a human-readable
report. A test case on a birdhouse model
demonstrates that the system correctly diagnoses overhangs, selects
appropriate defect mitigation strategies, and proposes physically valid
corrections, partially solving the geometry-to-language translation problem
central to LLM-driven CAD modification.
\end{abstract}

\section{Introduction}

Fused Deposition Modeling (FDM) is an additive manufacturing process which
builds parts layer by layer through depositing material along a sequence of
two-dimensional cross-sections \cite{rajan_fused_2022, pak_additivellm2_2026,
jadhav_llm-3d_2025}. Since each new layer must rest upon the one beneath it,
downward-facing surfaces tilted beyond roughly 45\textdegree{} from vertical
(overhangs) are difficult to print reliably without supports or geometric
modification. Design for Additive Manufacturing (DFAM) ~\cite{thompson2016dfam,
jadhav_llm-3d_2025} addresses these limitations by reshaping geometry to improve
dimensional accuracy, mechanical strength, and print efficiency while minimizing
post-processing. Modern slicing software can flag problematic regions and
generate support structures, but be corrective decisions must be made by the
engineer, who must iterate manually between CAD and slicer platforms until the
part is suitable for printing.

A small but rapidly growing body of work has begun applying language and
vision-language models at the CAD interface, organized in recent surveys along
generation, modification, and analysis axes \cite{llm_cad_survey1,
llm_cad_survey2}. On the generation side, Text2CAD produces sequential CAD
designs from natural-language prompts spanning beginner-to-expert detail
\cite{khan2024text2cad}, Query2CAD generates parametric models directly from
natural-language queries using FreeCAD \cite{query2cad}, and CADSmith
\cite{barkley_cadsmith_2026} utilizes CADQuery based Python bindings to generate
parametric models. ChatCAD extends this idea to multimodal LLM-guided CAD
drawing restoration in zero-shot settings \cite{chatcad}, and CADCodeVerify uses
vision-language models to validate generated CAD code \cite{cadcodeverify}. In
additive manufacturing specifically, large language models have been applied to
the domain of defect prediction in various build monitoring tasks
\cite{additivellm, pak_additivellm2_2026, jadhav_llm-3d_2025}, complementing
real-time vision-based defect-detection systems for FFF
printing~\cite{fff_defect}. These systems share a common pattern in that
language or images map to new geometry or to defect classes, but no system to
the authors' knowledge modifies an existing part's underlying B-Rep for more
favorable manufacturing conditions.

Complementary literature on Automatic Feature Recognition (AFR) provides the
geometric understanding needed to fill that gap \cite{shah_discourse_2000}.
Early work used hand-engineered rules and shallow neural networks to map
manufacturing features from CAD geometry \cite{afr_rules,afr_nn}. The
deep-learning era brought volumetric approaches: 3D ShapeNets pioneered learned
representations of 3D shape from voxel grids \cite{shapenets}, and subsequent 3D
CNNs added gradient-based visual explanations for machining feature recognition
\cite{cnn3d_machining}. More recent work returns to the native B-Rep: UV-Net
learns directly from parametric surfaces and curves \cite{jayaraman2021uvnet},
while Hierarchical CADNet extracts machining features from B-Rep face-adjacency
graphs using hierarchical message passing trained on the MFCAD++ corpus
\cite{hierarchical_cadnet}. Hierarchical CADNet's graph formulation directly
inspired the Phase~5 GraphSAGE embedder. These geometric methods rest on
foundational machine-learning advances. Graph Convolutional Networks (GCNs)
introduced spectral graph convolutions for semi-supervised node classification
\cite{gcn}, and GraphSAGE generalized this to inductive learning on unseen nodes
via neighbor sampling and aggregation \cite{graphsage}, making it directly
applicable to \texttt{STEP} files the system has never encountered. PyTorch
Geometric provides the implementation substrate widely used for these methods
\cite{pyg}. For visual verification, the Vision Transformer demonstrated that
pure-attention architectures can match or exceed convolutional networks at image
recognition \cite{vit}, underwriting the multimodal capability exploited in the
GPT-4o verifier ~\cite{openai2024gpt4o}.

These threads converge on a clear gap. Existing approaches address isolated
sub-problems but stop short of closed-loop DFM repair: Text-to-CAD systems
generate geometry from natural language yet cannot reason over or modify an
existing part~\cite{wu2021deepcad, khan2024text2cad}; B-Rep feature-recognition
networks classify face types with high accuracy but produce no geometric
modifications in response~\cite{
wu2024aagnet}; and automated manufacturability-analysis tools predict
non-manufacturable regions without generating corrective
fixes~\cite{balu2017dfm, zhong2025deepmill}. None of these lines closes the loop
where an existing CAD part is parsed, reasoned about, and modified back into a
manufacturable form. This presents a question: Can an agentic LLM system
automate Design for Manufacturing for FDM? The focus would be on the tractable
subset of automatically detecting overhangs and applying targeted DFM strategies
to resolve them.

This work introduces \textit{AgentsCAD} which takes in a \texttt{STEP} file with a manufacturing defect (e.g. overhang)
and produces (1) a modified STEP file with overhangs resolved and (2) a
human-readable report describing what was changed and why. The system is
primarily composed of four core features. First, a 
JSON serialization of geometry-to-language representation is used to translate B-Rep geometry,
topology, and shape descriptors into a compact LLM-readable prompt. Second, a
blackboard multi-agent architecture ~\cite{hayesroth1985blackboard} in which a shared-state orchestration
pattern makes every phase interchangeable without modifying any
other agent. Third, a hierarchical GraphSAGE feature embedder, a GNN trained on
MFCAD++ dataset ~\cite{hierarchical_cadnet} that produces 25-class semantic face labels for injection into the LLM
prompt, with a systematic comparison against a GCN baseline. Fourth, an
end-to-end working demonstration of validated test parts of
varying complexity, incorporating VLM-based visual verification and
human-in-the-loop prompt engineering. 

 \begin{figure}[H]
   \centering
   \includegraphics[width=0.75\linewidth]{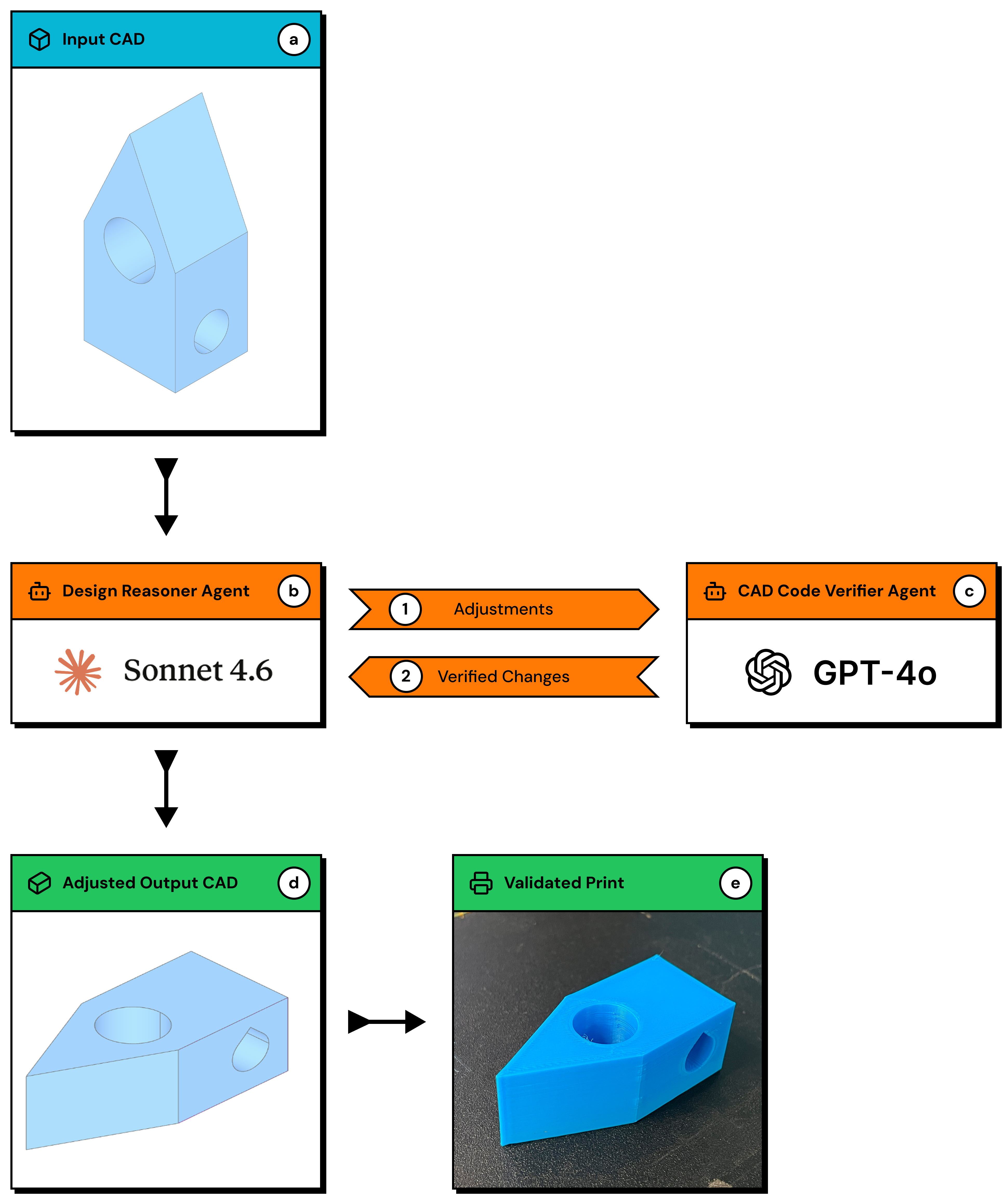}
    \caption{
        Agentic system for adapting CAD models for FDM 3D printing. Example (a)
        displays simplified bird house model with potential orientation and
        geometry adjustments. CAD model is delivered to the (b) Design Reasoner
        agent which provides suggestions and optimizations for adjustments which
        the (c) CAD Code Verifier Agent then evaluates. Iteration between the
        two agents proceed until a sufficient adjusted (d) output CAD model is
        generated and (e) validated via FDM 3D printing.
    }
   \label{fig:main}
 \end{figure}

\section{Related Works}

LLM-3D Print by Jadhav et al. \cite{jadhav_llm-3d_2025} explores the use of an
agentic system for the purpose of FDM print optimization. In this work in-situ
monitoring techniques via optical camera are used to assess the quality of a
print after each layer to detect potential FDM defects \cite{jadhav_llm-3d_2025,
hu_real-time_2024}. Unlike other related in-situ based monitoring techniques
which utilize specialized models to detect specific features
\cite{pak_thermopore_2024, george_beadsight_2024, bostan_accurate_2025,
wang_image2gcode_2026, ogoke_deep_2024, ogoke_deep_2025,
ogoke_convolutional_2023}, LLM3D print utilizes the base ChatGPT 4o to analyze
the obtained images \cite{jadhav_llm-3d_2025}. Utilizing custom firmware, the
system is capable of executing real-time parameter changes through tool calling
to repair builds optimizing on quality and defect reduction in the following
layers \cite{jadhav_llm-3d_2025}. The results of the agentic system were
evaluated by a cohert of qualified engineers and mechanically validated with
compression testing \cite{jadhav_llm-3d_2025}. Additionally LLM3D print was
evaluated on various multiple FDM printer setups, showcasing the agentic
system's versatility on any commercial printer platform
\cite{jadhav_llm-3d_2025}.

In Agentic Additive Manufacturing Alloy Evaluation by Pak et al.
\cite{pak_agentic_2026}, the authors develop an agentic system for the search
and evaluation of alloys suitable for fabrication via additive manufacturing.
The developed system leverages the Model Context Protocol (MCP)
\cite{anthropic_introducing_nodate} allowing for universal compatibility of the
developed tools to various large language models. The system is composed of 3
main subagents including the workspace agent primarily involved with managing
the output and folders of the two other agents those being the ThermoCalc
subagent and the Additive Manufacturing subagent \cite{pak_agentic_2026}. The
ThermoCalc subagent utilizes TC Python in order to generate the material
parameters such as solidus and liquidius temperatures and specific heat capacity
values from an arbitrary elemental composition \cite{pak_agentic_2026}. The
Additive Manufacturing subagent is responsible for generating the process map of
the given material parameters and a range of power and velocity combinations
using Rosenthal's approximation of a melt pool \cite{rosenthal_theory_2022,
pak_agentic_2026}. From this work, it was shown that an agentic system is
capable of generating preliminary lack of fusion process maps for various alloy
compositions utilized for specific applications \cite{pak_agentic_2026}.

RocketSmith \cite{pak_rocketsmith_2026} investigates the use of an agentic
system to aid in the design and manufacturing of high powered rockets. This
system combines the use of various tools including OpenRocket, CADSmith
\cite{barkley_cadsmith_2026}, and PrusaSlicer to optimize over flight parameters
such as stability, generate the associated parametric \texttt{STEP} files, and
output the manufacturing designs and files respectively. The system provides a
graphical user interface allowing human-in-the-loop guidance and adjustment to
design along with a collection of subagents and skills to aid with workflow
specific tasks \cite{pak_rocketsmith_2026}. Four different rockets were
generated using the agentic system and flight tested achieving a maximum
altitude accuracy of 84\% to the calculations \cite{pak_rocketsmith_2026}. Two
of the four rockets were successfully recovered in re-flyable condition
showcasing the agentic system's capability to generate suitable designs for high
powered rockets.

\section{Background}
\subsection{GCN}

A Boundary Representation (B-Rep) solid encodes geometry as a graph whose
nodes are faces and whose edges connect topologically adjacent faces. This
native graph structure makes Graph Neural Networks (GNNs) the natural
embedding tool. The Graph Convolutional Network (GCN) of
\citet{gcn} computes node representations by spectral convolution
with a first-order localised filter. At each layer $l$ the update rule is
\begin{equation}
    \mathbf{H}^{(l+1)} = \sigma\!\left(
        \tilde{\mathbf{D}}^{-1/2}\tilde{\mathbf{A}}\tilde{\mathbf{D}}^{-1/2}
        \mathbf{H}^{(l)}\mathbf{W}^{(l)}\right),
    \label{eq:gcn}
\end{equation}
where $\tilde{\mathbf{A}} = \mathbf{A}+\mathbf{I}$ is the adjacency matrix
with self-loops, $\tilde{\mathbf{D}}$ is its degree matrix, and $\sigma$ is a
nonlinear activation.
\subsection{GraphSage}
GraphSAGE  \cite{graphsage} replaces GCNConv layers with SAGEConv at
both levels of the B-AAG, while preserving the same two-level
pooling structure, number of layers, hidden dimension, dropout,
and classification head. GraphSAGE aggregates neighbor
features separately and \textit{concatenates} the result with
the node's own current representation:
\begin{equation}
  h_v^{(l+1)} = \sigma\!\left(W^{(l)} \cdot
    \mathrm{CONCAT}\!\left(h_v^{(l)},\;
      \mathrm{MEAN}_{u \in \mathcal{N}(v)} h_u^{(l)}
    \right)\right).
\end{equation}
\subsection{RAG}
Large language models (LLMs) used in agentic systems lack persistent memory across invocations: each call begins with an empty context window, discarding
any nuance accumulated over prior parts. Retrieval-Augmented Generation
(RAG) \cite{lewis2020retrieval} addresses this by conditioning generation on documents retrieved from an external index at inference time. In the original
formulation, a dense retriever maps queries and documents into a shared vector space; at generation time, the $k$ nearest documents are prepended to the prompt, grounding the model's output in non-parametric, updatable knowledge.
\section{Methods}

\subsection{Representational Challenges}

Representing CAD geometry to an LLM is a non-trivial task
\cite{jadhav_large_2026, wang_image2gcode_2026, barkley_cadsmith_2026}. A
\texttt{STEP} file encodes thousands of faces, edges, and vertices in a binary
B-Rep format that language models cannot parse directly. The initial hypothesis
proposed that raw geometric scalars alone (surface normals, areas, centroids)
would provide sufficient context for an LLM to reason about manufacturability.
In practice this approach produced a representation that was simultaneously too
verbose and too unstructured: the model received large arrays of floating-point
numbers with no indication of what those values mean in a manufacturing context,
and spatial relationships between adjacent faces were entirely absent. A face
with a downward normal adjacent to a cylindrical surface is almost certainly a
through-hole, but without a semantic label the model must infer this solely from
numerical primitives. Equally, no single fix resolves every overhang. The same
downward-facing face may call for a fillet, a reorientation, or a part split
depending on surrounding context.

CadQuery provided a Python interface to traverse the B-Rep topology of a loaded
STEP file, exposing each face as a structured object with a semantic
surface-type label: Planar, Cylinder, Cone, or BSpline alongside area, outward
normal, and centroid. After parsing, every face is described by features such as
its surface type, area, centroid, and signed tilt angles. The
\texttt{radius\_of\_gyration\_mm} indicates the approximate footprint on the
build surface, while the \texttt{elongation\_index} captures how lopsided that
footprint is, allowing faces of similar area but different shape to be
distinguished. At the part level, a total bounding box and centre of mass anchor
the spatial reasoning. Where higher-fidelity analysis was required, the
underlying OpenCASCADE (OCCT) kernel was called directly: surface-property
routines extract the inertia tensor per face, from which radius of gyration and
elongation index are derived as stability proxies; parametric UV-space sampling
via BRepAdaptor surfaces captures curvature variation across each face. Equally
important, OCCT's edge-to-face topology traversal maps shared-edge adjacency
between faces, encoding the spatial relationships that a flat scalar array
cannot represent. Together these two layers produce a structured intermediate
representation. Semantic type labels, orientation metrics, and an explicit face
adjacency graph. This is legible to a language model where the raw binary
\texttt{STEP} format is not, and that carries enough topological context to
support inferences of the kind illustrated above: a planar face sharing an edge
with a cylindrical face is a candidate through-hole; absent that adjacency, the
model has only a normal vector and an area to reason from. Any usable
representation must therefore \textit{preserve enough spatial and topological
context to make a contextual decision possible} while remaining compact enough
to fit inside a model's context window. The complete prompt is a single
\texttt{JSON} object containing \texttt{features}, \texttt{links}, and
\texttt{overhangs} sections. This representation provides a sufficient amount of
context for the LLM to reason over individual faces by ID while remaining
compact enough to leave headroom for CoT output and tool-use turns. Lastly, With
GraphSAGE weights present, each face is augmented using a \texttt{feature\_type}
label (e.g.\ \texttt{rect\_pocket}, \texttt{chamfer}) and a
\texttt{feature\_confidence} score, and these fields are folded into both the
per-face record and the adjacency descriptions. The topology graph is serialized
into a \texttt{links} list of \texttt{adjacent\_to} relations between face IDs.

\subsection{System Architecture}
\label{sec:pipeline}
\subsubsection{Blackboard Architecture for Multi-Agent Coordination}
\label{sec:blackboard}

Multi-agent LLM frameworks such as LangChain and CrewAI impose a
conversational or role-based execution model in which agents communicate via
message passing, making it difficult to enforce strict phase ordering, inject
deterministic rule-based stages, or swap individual models without refactoring
the coordination logic. AgentsCAD instead organises its pipeline around the
\textit{blackboard pattern} \cite{hayesroth1985blackboard}, a classical
architecture in which a single shared-state object is read and written by
independent agents in a fixed phase order. This design decouples every phase
from every other: modifying the geometry parser requires changing only that
agent and one configuration entry, leaving the downstream reasoner and
verifier untouched. Three further properties motivated this choice.
\textit{Deterministic floor}: initial defect identification is entirely
rule-based, flagging any face whose surface normal satisfies
$\theta = \arccos(\hat{n}_\mathrm{face}\cdot\hat{z}_\mathrm{plane}) \leq 0$
as a potential overhang without invoking an LLM. This establishes a
verifiable ground truth that the language model never has to reproduce and
cannot corrupt. \textit{State-reactivity}: agents read the blackboard between
phases and conditionally trigger or skip downstream work based on prior
output; if the reasoner finds no printability concerns, the modifier and
verifier phases are skipped entirely, avoiding unnecessary API calls.
\textit{Dual-model deployment}: the reasoning phase is handled by Claude
Sonnet 4.6 using chain-of-thought prompting \cite{wei2022chain}, while a
separate GPT-4o instance serves as the vision-language verifier; the
blackboard pattern makes it trivial to replace either model independently,
as neither agent has knowledge of the other's identity.

\subsubsection{Agentic System}
\label{sec:agentic system}
The system is carried out through nine phases with two conditional
sub-phases (Figure~\ref{fig:main}).

%
The system first imports the \texttt{STEP} file and extracts several scalar
attributes per face which include: surface type, area, centroid coordinates,
axis signed tilt angles, radius of gyration, and a unitless elongation index
distinguishing dense from needle-like geometry. Bounding box and center of mass
are written to a shared blackboard for downstream access. From there, every face
normal is compared against the build plane's z-axis using the arccosine of their
dot product, and any face at or below zero degrees is flagged as an overhang.
The part is then rendered as either a static PNG or an interactive 3D
environment for human inspection, establishing the visual baseline the verifier
will use later. Shared-edge adjacencies are resolved next and written to the
blackboard as a JSON adjacency structure. A GraphSAGE model pretrained on 59,665
parts then produces 128-dimensional embeddings and 25-class semantic feature
predictions per face; when pretrained weights are unavailable, node2vec embeddings are substituted, which are retained locally but withheld from
the language model so the system can continue on geometry and topology alone.
With the blackboard fully populated, the assembled \texttt{JSON} prompt is
passed to the design reasoning module, which applies chain-of-thought (CoT)
\cite{wei_chain--thought_2023} reasoning grounded by MCP tools to recommend
geometric changes. A scripted orientation sub-phase evaluates up to four
candidate rotations and commits to the best one before any modification is made.
The recommended operations (fillets, chamfers, teardrops, extrusions) are then
applied programmatically, and the resulting geometry is rendered and inspected
by a vision-language model to verify the changes visually. The system workflow
concludes by writing the optimized \texttt{STEP} file, a verification render, a
plain-text recommendations file, and a full blackboard snapshot for human
review.

\subsection{Evolving Geometric Understanding:
            GCN to GraphSAGE to RAG}
\label{sec:gnn}

\subsubsection{Motivation} Raw scalar geometry exposed a fundamental
tension. The LLM received large arrays of numbers (normals,
areas, tilt angles) but it had no indication of what those primitives meant in a manufacturing context nor did it capture inter-face geometric relationships in the scalar representation.
Pockets, slots, and steps are defined by \textit{how faces relate
to one another}, not by any single face in isolation. This
motivated injecting learned semantic feature labels via a graph
neural network trained on MFCAD++.

\subsubsection{Baseline: Hierarchical GCN}~\cite{gcn} GCN is applied hierarchically across the two-level Bipartite
Attributed Adjacency Graph (B-AAG) structure of MFCAD++,
stacking three GCNConv layers (hidden dim~64, dropout~0.5) at
each level. GCN is an appropriate baseline because the B-Rep
face-adjacency graph is the native data format and
\textit{homophily} holds strongly: faces belonging to the same
machining feature are topologically adjacent by construction.
However, GCN is \textit{transductive}, it learns node
embeddings tied to a fixed training graph and cannot generalize
to unseen graphs without retraining. In MFCAD++, each of the 59{,}665 parts is a separate graph of different sizes and
topology, which makes transductive inference incompatible with the
deployment scenario. Additionally, GCN's symmetric
normalization dilutes each face's own geometric attributes into
the neighborhood aggregate, damaging recognition of minority
classes whose distinctive local geometry is the primary
distinguishing signal.

\subsubsection{Variant: Hierarchical GraphSAGE}

The concatenation structure of GraphSAGE has two consequences that are critical for
CAD-geometry inference.  First, \emph{inductiveness}: because GraphSAGE
learns an aggregation function rather than per-node embeddings, the trained
model can produce embeddings for faces on entirely new parts at test time.
This is essential for AgentsCAD, where the model is expected to classify faces
on user-supplied STEP files it has never encountered.  Second,
\emph{self-feature preservation}: each face's own geometric attributes
(surface type, area, centroid, tilt angles, radius of gyration) are carried
forward as a direct input to every layer, ensuring they are never overwritten
by neighbourhood averaging. This is particularly valuable for rare feature
types such as \emph{triangular through slot} or \emph{slanted through step},
whose primary distinguishing signal is their local geometry rather than their
graph neighbourhood. AgentsCAD applies GraphSAGE hierarchically across the
two-level Bipartite Attributed Adjacency Graph (B-AAG) structure of MFCAD++,
stacking three SAGEConv layers (hidden dimension 64, dropout 0.5) at each
level. Face-level input features are optionally augmented with UV-sampled
surface normals following UV-Net \cite{jayaraman2021uvnet}, adding mean
normal, normal standard deviation, and a UV-coverage ratio to the base
five-dimensional per-face descriptor.

\subsubsection{Feature augmentation: UV-net surface normals}
Beyond the architectural variant, this work evaluates augmenting the
face-level node features~$V_1$ with UV-sampled surface normals
from the underlying B-Rep geometry, following
UV-Net~\cite{jayaraman2021uvnet}. For each face a $5{\times}5$
UV grid (retaining only points that pass the trim-boundary
test) is sampled followed by the computation of three additional descriptors: mean surface
normal (3D), normal standard deviation (3D), and UV coverage
ratio (1D). These 7 descriptors are concatenated with the
existing 5-dimensional $V_1$ features, producing a
12-dimensional face input. This augmentation applies to both
GCN and GraphSAGE, allowing independent evaluation of the architectural
and feature-level contributions independently.

\subsubsection{RAG memory} Even with learned
semantic labels, the reasoning agent lacks memory across parts. In AgentsCAD, RAG serves as a \emph{cross-part memory layer}. After each
analysis, the blackboard state (geometry descriptors, feature predictions,
and the LLM's reasoning trace) is encoded as a 256-dimensional context vector
and indexed in a FAISS store.  When analysing a new part, the $k$ most
similar past decisions are retrieved and prepended to the reasoning prompt. This allows the system to accumulate technical nuance over time without
retraining the underlying GNN or fine-tuning the LLM. For
example, learning that teardrop modifications are unreliable on
holes wider than 50~mm, or that certain feature combinations
reliably call for a split rather than a fillet. This enables
the system to improve steadily as it processes more parts,
without retraining the underlying model. The RAG layer therefore converts agentsCAD from a stateless
classifier into a system that improves monotonically as it processes more
parts.
\subsection{Design Reasoning Agent}
\label{sec:reasoner}

The Design Reasoner is a Claude~Sonnet~4.6 agent provided with
the assembled JSON geometry prompt and a comprehensive
manufacturing system prompt encoding DFM heuristics: angle
conventions, overhang severity thresholds, stability criteria,
and modification trade-offs. The agent uses CoT reasoning to
analyze the model for overhangs, print stability, stress
concentrations, and thin features, following a mandatory
decision order. First, reorientation is considered globally: the agent calls \texttt{check\_orientation\_overhangs} to ground the proposed rotation in computed geometry, then calls \texttt{lay\_face\_to\_build\_surface} to compute exact $x$/$y$ rotation angles for a target face. Second, local modifications, such as filet, chamfer, and teardrop, are applied after the orientation is fixed. Third, support structures are flagged when no geometric fix suffices. Fourth, splitting the part is used as a last resort and overrides all other actions when invoked.

The two MCP tools ground the agent's ``mental rotation'' in
real geometry, preventing the coordinate-transform
hallucinations that arise when an LLM must reason about 3-D
rotations from text alone. The agent returns a typed JSON
recommendations dictionary consumed directly by Phase~7.

\subsection{Part Modification}

The JSON modification recommendations are assigned by target
\texttt{face\_id} and programmatically executed via CadQuery.
The reorient operation is always performed \textit{last}, because
reinterpreting the STEP file after rotation reallocates
\texttt{face\_id} values, which would invalidate any earlier
face-specific operations.

\subsection{Visual Verification}
\label{sec:vlm}

The verifier is a novel adaptation of
CADCodeVerify~\cite{cadcodeverify}, originally proposed for
validating generated CAD code, repurposed here for geometry
modification. After the modifier applies a set of operations,
the part is rendered from four angles into a $2{\times}2$
composite at $2048{\times}2048$ pixels with edges visible. The
system then generates one targeted yes/no question per
applied modification (e.g.\ for a fillet,
\textit{``Do edges of \texttt{face\_X} appear smoothly rounded
(${\sim}R$~mm)?''}) together with general integrity and
overhang questions that always run, catching silent modifier
failures even when no operation was applied. GPT-4o replies
Yes/No/Unclear with a chain-of-thought rationale; failures are
logged to the blackboard.

\subsection{Human-in-the-Loop Validation}

Beyond automated VLM verification, the development of a solidified workflow relied heavily on a
\textit{prompt-to-structure} approach in which the authors'
manufacturing expertise was used to design the system prompt
and MCP tool schemas that scaffold the LLM's reasoning toward
correct DFM logic. This process was iterative: each test case
exposed gaps in the agent's reasoning that were closed by
tightening angle conventions, adding mandatory tool-call
sequences, or expanding the modification action space.
Structured prompt engineering proved as important as any
architectural choice: an LLM without geometric grounding
consistently hallucinated rotations, while the same model with
the \texttt{check\_orientation\_overhangs} tool and explicit
angle invariants produced reliable, physically valid plans.

\section{Results and Discussion}

\subsection{Graph Embedder: GCN vs.\ GraphSAGE}

Table~\ref{tab:gnn_results} reports test-set performance across
all model configurations on the MFCAD++ held-out split. Macro
F1 is the primary metric: it computes F1 independently for each
of the 25 classes and averages without weighting by class
frequency, appropriately penalizing failures on rare feature
types.

\begin{table}[h]
  \caption{Test-set performance across all model configurations
           on MFCAD++. Macro~F1 is the primary metric.
           Best result per column in \textbf{bold}.}
  \label{tab:gnn_results}
  \centering
  \begin{tabular}{llcccc}
    \toprule
    Architecture & Features
      & Accuracy & Macro F1 & Macro Prec. & Macro Rec. \\
    \midrule
    GCN (baseline) & $V_1$ only  & 0.443 & 0.338 & 0.336 & 0.396 \\
    GCN            & $V_1{+}V_2$ & 0.363 & 0.306 & 0.345 & 0.377 \\
    GCN            & UV-net      & 0.538 & 0.469 & 0.472 & 0.552 \\
    GraphSAGE      & $V_1$ only  & 0.638 & 0.545 & 0.534 & 0.579 \\
    GraphSAGE      & $V_1{+}V_2$ & 0.794 & 0.727 & 0.716 & 0.753 \\
    GraphSAGE      & UV-net
                   & \textbf{0.850} & \textbf{0.785}
                   & \textbf{0.774} & \textbf{0.805} \\
    \bottomrule
  \end{tabular}
\end{table}

\subsubsection{GraphSAGE vs.\ GCN} GraphSAGE outperforms GCN
across all three feature representations, confirming the
primary hypothesis. On $V_1$-only features, GraphSAGE
achieves a macro F1 of 0.545 versus GCN's 0.338, a 61\%
relative improvement from the architectural change alone.
With UV-net augmentation, GraphSAGE reaches 0.785~F1 and
85.0\% accuracy versus GCN's 0.469~F1 and 53.8\% accuracy.
The performance gap widens as feature richness increases,
suggesting that GraphSAGE's concatenation-based aggregation is
better equipped to exploit additional information than GCN's
normalized averaging.

\subsubsection{Effect of hierarchical aggregation} The
$V_1 \rightarrow V_1{+}V_2$ transition has starkly different
effects on the two architectures. For GraphSAGE, adding the
facet-level pass yields the largest single improvement of any
step: F1 increases from 0.545 to 0.727~(+0.182). For GCN,
the same addition \textit{decreases} performance: F1 drops from
0.338 to 0.306. This counterintuitive result is consistent
with GCN's symmetric normalization, which merges pooled facet
context into the face representation via a weighted sum,
diluting the face's own geometry. GraphSAGE's explicit
concatenation preserves self-features, allowing the richer
hierarchical signal to help rather than hurt.

\subsubsection{MeshViz: Prediction visualization on B-Rep Geometry} 
\begin{figure}[H]
  \centering
  \begin{subfigure}[t]{0.44\linewidth}
    \centering
    \includegraphics[width=\linewidth]{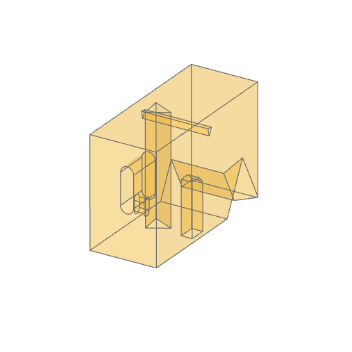}
    \caption{Original Part}
  \end{subfigure}\hfill
  \begin{subfigure}[t]{0.49\linewidth}
    \centering
    \includegraphics[width=\linewidth]{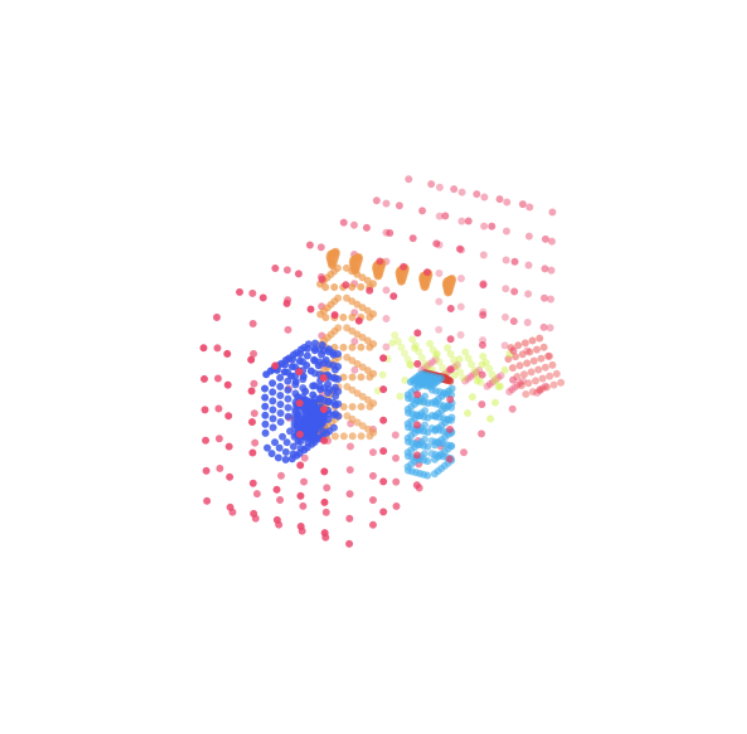}
    \caption{Predicted (97\% face acc.)}
  \end{subfigure}
  \caption{Per-face feature classification visualization for a simple part (GraphSAGE+UV-net). Predicted faces are colored by class; red faces are misclassified.}
  \label{fig:meshviz}
\end{figure}

MeshViz is a post-training visualization tool that renders model predictions directly onto B-Rep face geometry. It is used to ground the quantitative results in geometric intuition. After training concludes, MeshViz loads the STEP file for each sampled test part via CadQuery, runs inference on the corresponding held-out test graph, and tessellates each face's 3D surface by sampling an 8×8 UV grid over the parametric domain via the OpenCASCADE kernel, discarding points that fall outside the face's trim boundary. Each face's resulting point cloud is colored by prediction outcome: the face's assigned class color if the prediction matches the ground-truth label, and red otherwise. The colored point clouds are logged to W\&B as interactive Object3D objects for real-time inspection during experiment tracking, then rendered as static figures for this paper. Figure~\ref{fig:meshviz} shows representative samples from the best model (GraphSAGE UV-net).

The visualizations reveal that misclassified faces cluster at feature boundaries (the transition zones between a machining feature and the surrounding stock material). Interior faces of well-defined features are almost universally correctly classified, while boundary faces are consistently confused with adjacent feature types or the base stock. This is geometrically interpretable: boundary faces have mixed neighborhoods containing both feature-internal and feature-external adjacency, producing ambiguous aggregated representations regardless of the aggregation operator used. This finding suggests that boundary-aware loss weighting or explicit boundary detection as an auxiliary task could be a productive direction for future work.

\subsection{Test Case Validation}

The end-to-end functionality was validated on two test parts of
differing complexity. The tests were aimed to make sure the system could make consistent modification operations, provide consistent independent reasoning as well verify print stability. All renderings show the original part and orientation as well as the finished product following the system implementation.

\subsubsection{Birdhouse}

The birdhouse is a nine-face solid combining seven planar faces with two
cylindrical bores (\texttt{face\_4} and \texttt{face\_8}). In the default Z-up orientation both bores presented as severe overhangs
(tilt $-90$\textdegree), flagged as actionable by the rule-based overhang
detector. Prior to LLM reasoning, the RAG pipeline retrieved domain knowledge on
horizontal bores and teardrop compensation, injecting it directly into the
prompt and priming the reasoner with established FDM heuristics before any
tool calls were made.
The design reasoner issued six tool calls to evaluate candidate orientations,
calling \texttt{check\_orientation\_overhangs} to probe $90$\textdegree
rotations about each axis and \texttt{lay\_face\_to\_build\_surface} to
obtain exact Euler angles for three candidate bed faces before settling on a
recommendation.
The StrategyCouncil overrode the LLM's reorient target in favour of the
bore-side-lay heuristic: laying \texttt{face\_6} ($1{,}463$~mm$^2$) flat
on the bed at $X=90$\textdegree aligned the bore axis of \texttt{face\_4}
with the build direction, eliminating it entirely.
One bore (\texttt{face\_8}) remained horizontal after reorientation.
The council correctly rejected two spurious chamfer candidates, one on the
print-bed face and one on a vertical wall, before unanimously selecting a
teardrop modification for the remaining bore.
The modifier applied the teardrop to \texttt{face\_8}, expanding the face
count from 9 to 12, and a post-modification overhang recheck confirmed
zero actionable overhangs on the final geometry.
OCCT solid validation passed with a volume delta of $-0.75\%$, consistent
with the small material removal expected from a teardrop profile.
The VLM verifier returned two inconclusive flags on wall thickness and
cross-sectional area; both were attributable to insufficient render
resolution rather than genuine geometry defects, as the verifier's own
reasoning acknowledged the modifications were consistent with the intended
teardrop action and that the renders did not provide sufficient detail for
a definitive assessment.
The complete resolution sequence, two initial overhangs, one eliminated by
reorientation, one resolved by teardrop modification, was achieved in a
single pipeline iteration.


\begin{figure}[H]
  \centering
  \begin{subfigure}[b]{0.48\linewidth}
    \centering
    \includegraphics[width=\linewidth]{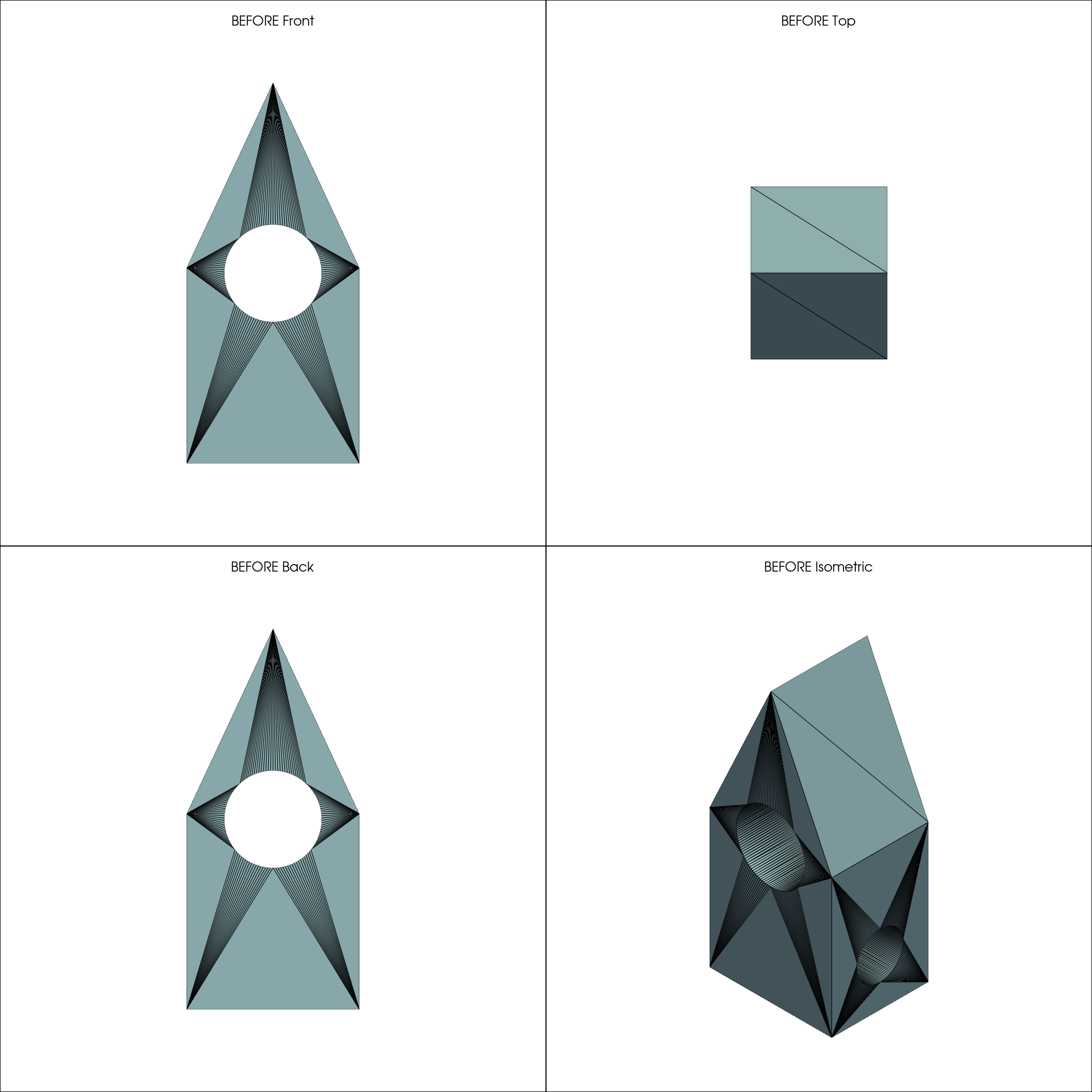}
    \caption{Original birdhouse CAD input}
    \label{fig:birdhousebefore}
  \end{subfigure}
  \hfill
  \begin{subfigure}[b]{0.48\linewidth}
    \centering
    \includegraphics[width=\linewidth]{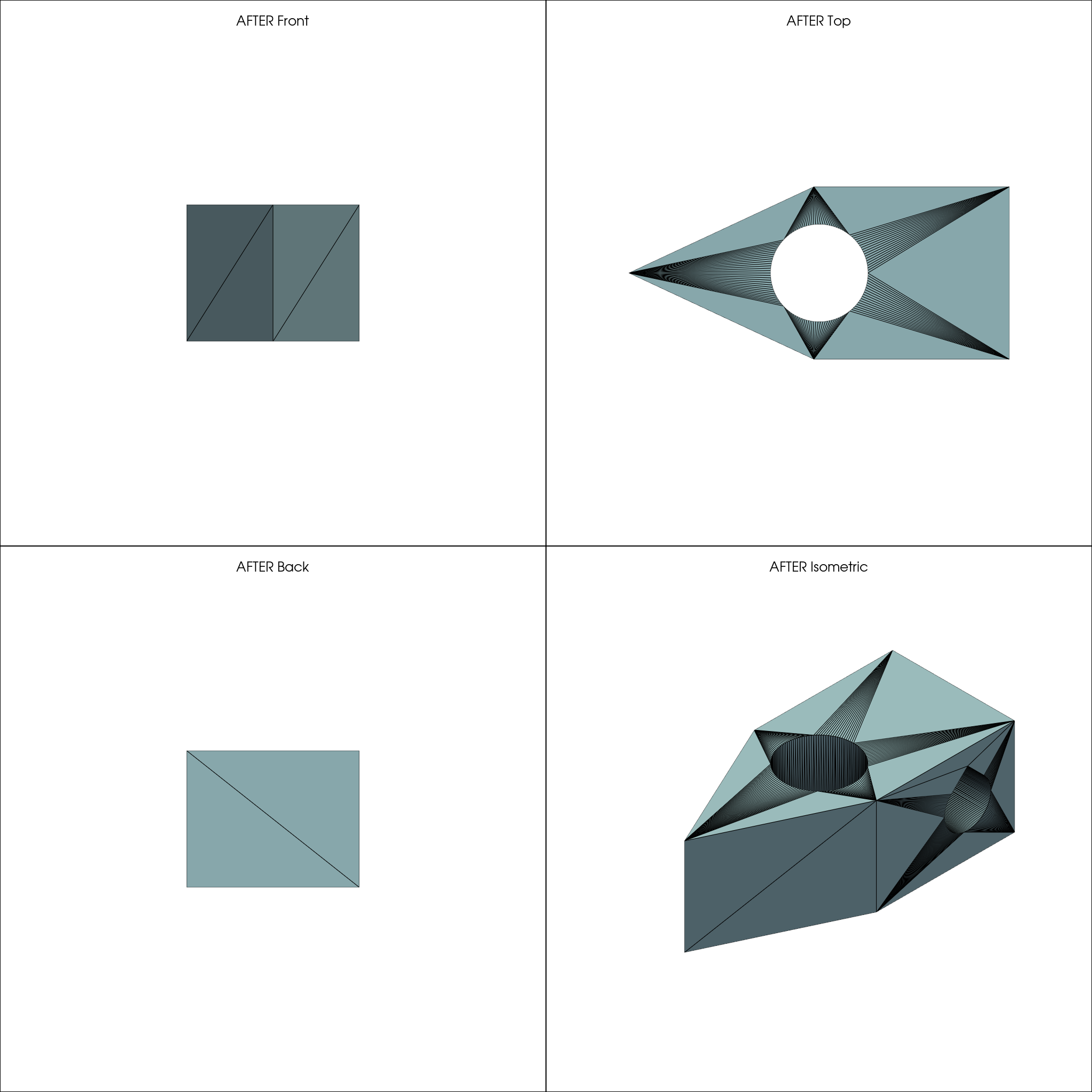}
    \caption{Modified birdhouse CAD output}
    \label{fig:birdhouseafter}
  \end{subfigure}
  \caption{Birdhouse test case before and after automated DFM modifications.}
  \label{fig:birdhouse}
\end{figure}

\subsection{Ablation: MCP Grounding Tools}

A key design question was whether tool-use grounding is necessary or whether the
LLM can reason about 3-D rotations from the \texttt{JSON} representation alone.
In early runs without the \texttt{check\_orientation\_overhangs} and
\texttt{lay\_face\_to\_build\_surface} MCP tools, the agent consistently
hallucinated rotation angles, proposing, for example, a $+45$\textdegree rotation
for a face whose correct reorientation requires $-30$\textdegree, then asserting
with high confidence that the result was overhang-free. With both tools enabled,
coordinate-transform errors were eliminated across all test cases: the agent
first queries the tool for the required rotation, then verifies the
post-rotation overhang map before committing. This ablation confirms that
geometric grounding via MCP tools is \textit{not} an optional enhancement but a
prerequisite for reliable DFM reasoning.

\section{Conclusion}

AgentsCAD demonstrates that raw \texttt{STEP} files can be turned into
DFM-improved parts by an agentic, blackboard-orchestrated system combining a
deterministic geometry parser, a Claude~Sonnet design-reasoner with a small
library of grounding tools, and a GPT-4o visual verifier. The system bridges the
gap between B-Rep geometry and LLM reasoning for the targeted problem of
overhang resolution, producing physically valid modified \texttt{STEP} files and
emitting human-readable manufacturing reports.

Three findings stand out. First, structured \texttt{JSON} geometry with explicit
topology instead of raw scalar arrays proves to be sufficient for an LLM to
perform contextual DFM reasoning, including stability analysis that goes beyond
overhang detection. Second, GraphSAGE's inductive, self-feature-preserving
aggregation outperforms GCN across all feature sets, with the largest gains on
richer representations where GCN's normalized averaging actively hurts
minority-class recognition. Third, MCP tool grounding is not an enhancement but
a prerequisite: without it, the agent hallucinates coordinate transforms with
high confidence.

\section{Future Work}

Larger parts incur both context-size pressure and ``lost-in-the-middle'' effects
as face counts grow, and the current prompt does not decimate or summarize
sub-graphs. The system uses a single agent for all DFM facets; decomposing
reasoning into dedicated sub-agents (e.g.\ a stress-concentration specialist)
would improve coverage. The system currently accepts only single parts with no
assembly support.

Four immediate extensions emerge from this work: a \texttt{SplitAgent} that partitions parts at LLM-specified planes; active GraphSAGE injection replacing prompt-based geometric primitives with topology embeddings consumed directly by the reasoner; expansion of the RAG memory to accumulate technical nuance across a growing library of parts; and defect-class expansion to robustly detect bridges, thin features, and internal bores. On a longer horizon, GPT-5's demonstrated capability for generating native CAD representations and interpreting open-ended user modification requests positions it as a natural upgrade for the Design Reasoner, possibly enabling the system to move beyond structured JSON prompts toward conversational, multimodal DFM collaboration with the full expressiveness of natural language.

\clearpage
\appendix

\appendix

\renewcommand{\thesection}{Appendix \Alph{section}}

\bibliography{references, peter_references}

\end{document}